# Personalised Pricing: The Demise of the Fixed Price?

**Joost Poort and Frederik Zuiderveen Borgesius**

**Abstract**

An online seller or platform is technically able to offer every consumer a different price for the same product, based on information it has about the customers. Such online price discrimination exacerbates concerns regarding the fairness and morality of price discrimination, and the possible need for regulation. In this chapter, we discuss the underlying basis of price discrimination in economic theory, and its popular perception. Our surveys show that consumers are critical and suspicious of online price discrimination. A majority consider it unacceptable and unfair, and are in favour of a ban. When stores apply online price discrimination, most consumers think they should be informed about it. We argue that the General Data Protection Regulation (GDPR) applies to the most controversial forms of online price discrimination, and not only requires companies to disclose their use of price discrimination, but also requires companies to ask customers for their prior consent. Industry practice, however, does not show any adoption of these two principles.

## 1    Introduction

On 15 May 2020, J.C. Penney filed for bankruptcy. This legendary department store founded in 1902 by John Cash Penney, with approximately 850 locations across the US,[1] had become famous for various retail innovations and a company policy it summarized as the 'golden rule': to treat others as you would like to be treated. One of the innovations J.C. Penney introduced was to offer the goods for low but fixed prices, at a time when haggling was the standard in stores. Is the bankruptcy of J.C. Penney just an example of the creative destruction that drives innovation and capitalism, in this case spurred by the Covid-19 crisis? Or does it signify the end of an era: the demise of the fixed price?

In our current data-driven world, an online seller or platform is technically able to offer every consumer a different price for the same product, based on information it has about the customers. For example, a seller can classify consumers according to their assumed wealth, or rather their price sensitivity, and charge those who are presumed to be less price sensitive higher prices. This chapter synthesises our previous findings on such online price discrimination,[2] a concept defined as differentiating the online price for identical products or services, based on the information a company holds about a customer.[3] We give examples of online price discrimination, discuss its underlying basis in economic theory, and its popular perception. The chapter argues that the General Data Protection Regulation (GDPR) applies to the most controversial forms of online price discrimination, and not only requires companies to disclose their use of price discrimination, but also the prior consent by the customer to it. Industry practice, however, does not show any adoption of these two principles – for reasons explored below. Price discrimination may also invoke the operation of anti-discrimination law (see Collado-Rodriguez/Kohl, Chapter 7),[4] competition law[5] or consumer protection law,[6] which are not further considered here.

## 2    Online price discrimination and the underlying economic theory

Many internet users will recognize the experience that after looking at shoes, a hotel room or a flight, they are followed around the web by advertisements for those products for weeks. This is an example of *behavioural targeting*. In this marketing technique, companies track consumers' online behaviour and use the information collected to display their targeted ads.[7] If companies can tailor ads to what they know about consumers, they can also tailor prices. 'Just as it's easy for customers to compare prices on the Internet, so is it easy for companies to track customers' behavior and adjust prices accordingly', consultants at McKinsey & Company wrote already in 2001, and added '[t]he Internet also allows companies to identify customers who are happy to pay a premium.'[8] Whether consumers are 'happy' to pay a premium is debatable; some evidence suggest otherwise. In an early incident, Amazon attracted criticism[9] when its customers, upon deleting the cookies from their browsers, would see the prices for DVDs drop. Amazon appeared to be charging higher prices to existing customers, or – put differently – appeared to be trying to attract new customers with lower prices. In response to the exposure, Amazon stated that it had tested different prices, but not adjusted prices to customer profiles based on their demographics,[10] and refunded affected customers.

The techniques of such discrimination have become more precise and innovative with time. In 2012, researchers evidenced online price discrimination by some US web stores based on their customers' neighbourhoods,[11] as revealed by the IP address of their computers. For example, the office supply store Staples charged lower prices if the website visitor was in a place where there was a competing brick-and-mortar store within a 20-mile radius. This pricing strategy makes sense from the seller's perspective, as most consumers will not drive an hour to save a few dollars on a box of printer paper. But if a consumer can buy printer paper in a shop just around the block, she will do so if the price in the online store is too high. This pricing strategy also had the unintended effect that, on average, customers on lower incomes paid higher prices. Similar research has also shown that a hotel room in Los Angeles can be 14% more expensive for someone with a Dutch IP address than for someone with an American IP address.[12] Online stores could also adjust prices based on the surfing behaviour of customers, as revealed by the cookies placed in their browsers.

If online price discrimination is possible, profitable and legally permissible, one would expect it to be pervasive. However, at this stage its prevalence appears to be limited,[13] for which there may be several explanations. First, some providers may refrain from price discrimination because they fear negative responses from consumers. Second, others may lack sufficient data or technology to adapt prices with sufficient precision, which would be remedied with easier and cheaper access to relevant technology. Finally, price discrimination may occur without in fact being noticed, as sellers can adapt prices in ways that are hard to detect, for example, when a provider advertises one price on its website but emails each consumer a different discount coupon. Many commentators expect that price discrimination will become more common as data-driven technology advances.[14]

While data-driven personalisation technology has provided new possibilities for price discrimination, it is not a new phenomenon. Classic examples are a lower conference rate for students than for corporate participants, or a reduced rate for children for theatre, cinema or flight tickets. According to underlying economic theory, three conditions must be fulfilled for price discrimination to work effectively: (i) the seller must be able to distinguish customers to know what price to charge to whom; (ii) the seller must have a sufficiently strong market position to be able to set prices above marginal costs; and (iii) resale must be impractical, costly, or prohibited in order to avoid arbitrage between customers.[15]

For online sales, these conditions are often met. Companies can distinguish between customers with increasing precision. Some providers, eg Amazon.com and Booking.com, have a strong market position and stimulate customer lock-in, which is likely to allow them to achieve higher margins. Finally, reselling is also often impossible (as, for example, in the case of plane tickets or hotel rooms), difficult or expensive. In combination with the fact that prices can be adjusted with ease and unnoticed, the online environment provides a highly fertile grounds for the proliferation of price discrimination.

As a matter of background, economists distinguish between first-, second- and third-degree price discrimination.[16] In *first-degree price discrimination,* each consumer is charged an individual price equal to his maximum willingness to pay. In practice, despite the rise of big data analytics, such an extreme form of price discrimination is still unlikely to occur, as sellers are unlikely to know each buyer's exact willingness to pay. Rather, first-degree price discrimination serves as a stylised benchmark to evaluate other pricing schemes. *Second-degree price discrimination* refers to situations where the price of a good or service depends on the *quantity* purchased, for example, through quantity discounts. In the cinema, for example, popcorn is often cheaper per gram if you buy a larger box. For second-degree price discrimination, the seller does not need information about the buyer, because buyers themselves choose the price by choosing the quantity. Loyalty programmes are also sometimes characterised as second-degree price discrimination, which is correct to the extent that such programmes involve a quantity discount over time: past purchases give a discount on future purchases.

By creating customer profiles, providers can also use loyalty programmes for personalised prices, which qualifies as third-degree price discrimination. In the case of *third-degree price discrimination*, prices differ between groups of buyers. Discounts for students, children, or the elderly are well-known examples. The distinction can also be geographical as, for example, when medicines and textbooks are sold at lower prices in developing countries. For third-degree price discrimination, it is not necessary to recognise individual buyers. Sellers only need to know the characteristics of the customer that are used to base the price differences on. Nevertheless, sellers often use unique identifiers to distinguish different types of buyers, such as a student card with a student number and photo or even a formal ID-card. This unique identification of customers helps to meet two of the most important conditions for price discrimination: differentiating between buyers and preventing arbitrage.

Online price discrimination usually works similarly. An online provider identifies a customer based on, for example, an IP address or login details, as a means to distinguish between groups of customers. However, compared to distinguishing students based on a student card, an online profile can be much more detailed and allow for much more sophisticated price discrimination. In this way, online third-degree price discrimination can shift towards the holy grail of perfect first-degree price discrimination, under which all consumer surplus is extracted for the benefit of the seller. For instance, an online provider could at the same time adjust prices based on past purchasing behaviour of a consumer depending on: whether she landed from a competitor's website; whether she lives in a poor or a rich neighbourhood (information which can be deduced from her IP address); whether she uses an Apple or a Windows device; and whether she uses a mobile device or a laptop.

Another pricing strategy widely used on the internet is *dynamic pricing* or *time-based pricing*. With dynamic pricing, a company (automatically) adjusts prices based on market conditions concerning supply and demand. For example, an airline will increase the price of tickets if a flight is almost fully booked. The airline will also charge higher prices on popular times and days, for example, for tickets to beach destinations during school holidays. Taxi platform Uber is also notorious for its dynamic prices.[17] Dynamic pricing is often mistaken for online price discrimination when, for example, the potential customer returns to the website and sees that the airline ticket has become more expensive. In this situation, the change in price may be due either to the departure date being closer or the aircraft being fuller (dynamic pricing), or the fact that the airline has seen that the customer looked at the ticket before and is therefore probably serious about a purchase (third-degree price discrimination).[18]

Economists are generally positive about price discrimination (and dynamic pricing). Price discrimination can benefit both buyers and sellers, leading to an increase of both consumer and producer welfare.[19] Price discrimination can help the seller to recoup his fixed costs without losing many potential customers and make a good or service accessible to buyers with a smaller purse, even if it will lead to higher prices for other customers. For the latter, however, price discrimination deprives them of consumer surplus. The more sophisticated the pricing scheme used by the seller, the more this will be the case, because a more accurate price discrimination scheme will extract more additional value from those willing to pay more. While this may improve efficiency and allow the seller to widen the scope of the market to marginal consumers, such benefits are 'financed' by the additional squeezing of higher prices from those willing to pay more. In addition, as the example of Staples charging higher prices to those far from competing retailers demonstrates, price discrimination may reinforce economic inequality by unintendedly leading to higher prices for poorer customers: not because they are poor, but as a by-product of other variables, such as the proximity of competing suppliers. Thus price discrimination can contribute to a cycle of economic inequality.

There is a large body of literature on the welfare effects of price discrimination under different assumptions about consumer demand, the information available to consumers and suppliers, etc. The impact of price discrimination on the average price of goods and services and thereby on total consumer welfare is not always clear.[20] The total welfare effects of consumers and producers combined are ambiguous too. Price discrimination may lead to a net welfare loss, when the loss by consumers who pay higher prices (based on price discrimination) is greater than the net gain by producers. And even sellers themselves may suffer a net welfare loss, if price discrimination intensifies competition for price sensitive consumers. Thus, the short answer to the question about the welfare effects of price discrimination is: it depends. To increase welfare, price discrimination must lead to a significant increase in total output by serving markets not previously served. However, even then, less price-sensitive consumers will often be worse off from price discrimination because they will, on average, pay higher prices.

The more price discrimination becomes personalised, the more welfare will generally shift from consumers to suppliers. The reason for this is that personalising prices more, entails placing consumers in smaller, more homogeneous segments and charging them a price that can be closer to their willingness to pay. For example, in the analogue world, prices may be differentiated between students and other consumers, based on the idea that the average student has less disposable income. However, within the student population, there will be poor students and rich students. In the digital world, suppliers may further differentiate within this student population, based on, for instance, the neighbourhood where they live and the type of device they use to navigate the internet. By doing so, sellers can charge richer students a premium.

## 3   Popular perceptions of price discrimination

Despite several sporadic pieces of evidence of online price discrimination in practice, it is largely unknown at what scale it is actually taking place. Nevertheless, it is relevant to know what the popular perceptions of price discrimination are, for instance to inform policy makers. This section provides empirical evidence regarding why personalisation might be concerning, especially as it expands its reach and power. To investigate consumer attitudes towards different forms of price discrimination and dynamic pricing, we conducted two surveys among a representative sample of the Dutch population of 18 years and older.[21]

### 3.1  Survey 1: general experiences and views

The first survey was conducted in April 2016 and had a response rate of 1233 (81.0%). The questions focused on consumers' experiences with online price discrimination and their attitudes towards it. In the questions, the term 'price discrimination' was not used because it could have a normative connotation. The survey described personalised pricing as follows:

> Web stores can adjust prices on the basis of data about an Internet user, such as the country where the user is based, or the time the user visits the web store. This makes it possible that two Internet users, who visit the same web store at the same time, see different prices for the same product.[22]

After this introduction, respondents were asked how often they had experienced this themselves. 57% said they had never experienced online price discrimination, while 4% said they experienced it often or very often. Around 40% asserted that they experienced online price discrimination rarely or occasionally.[23] Next, respondents were asked to indicate on a scale of 1 to 7 whether they think such practices should be prohibited. Figure 1 shows that a large majority is in favour of a ban: 72% choose 5, 6 or 7.

**Figure 1:** 'According to you, should such practices be prohibited?' (N=1233)

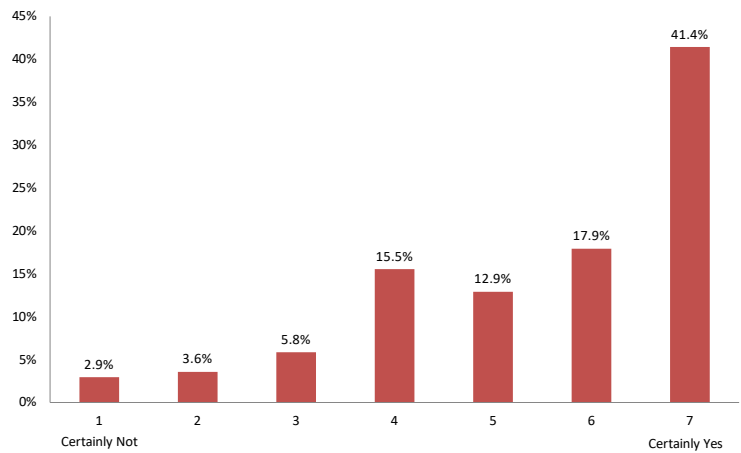

**Figure 2:** 'Would you find it acceptable if a web store gives a discount to you/others based on your/their online behaviour (such as the websites you/they have visited before)?' (N=1233)

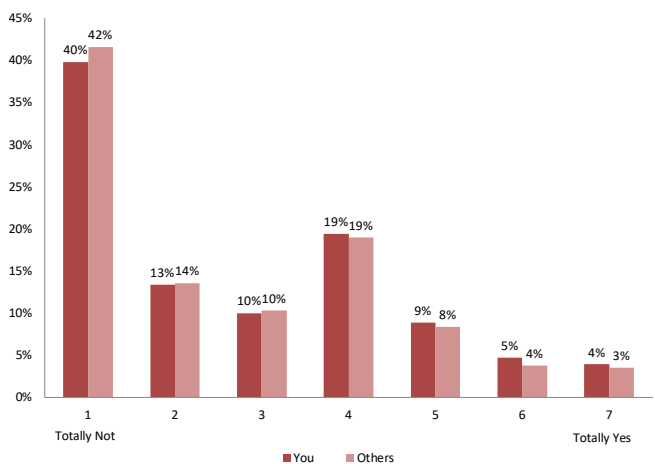

Asked whether online price discrimination is acceptable and fair, more than 80% indicate that they find it unacceptable and unfair to a certain extent, while only a few percent indicate they find it acceptable. The differences in outcomes between 'acceptable' and 'fair' are minimal, which suggests these concepts are closely linked in the perception of respondents. Figure 2 shows that if online price discrimination is framed as a personalised *discount*, acceptance increases slightly, but about 65% still find price discrimination unacceptable. Surprisingly, popular opinion hardly depends on whether the price discrimination favours themselves or others.

Finally, to find out *why* consumers do or do not agree with online price discrimination, we presented respondents with three propositions. Almost 80% agree with the statement that web stores should be obliged to inform customers about online price discrimination. And 56% are to some extent concerned about paying more than others, while 65% are worried about not noticing price adjustments.

### 3.2 Survey 2: specific examples

The second survey asked respondents for their opinion on specific examples. Fifteen examples were presented to them and, as in the first study, they were asked to indicate on a 7-point scale to what extent they considered them acceptable. This survey was conducted in November 2016 and had a response rate of 1202 (82.2%).

**Figure 3:** Net acceptability of different forms of price discrimination and dynamic pricing (N=1202)

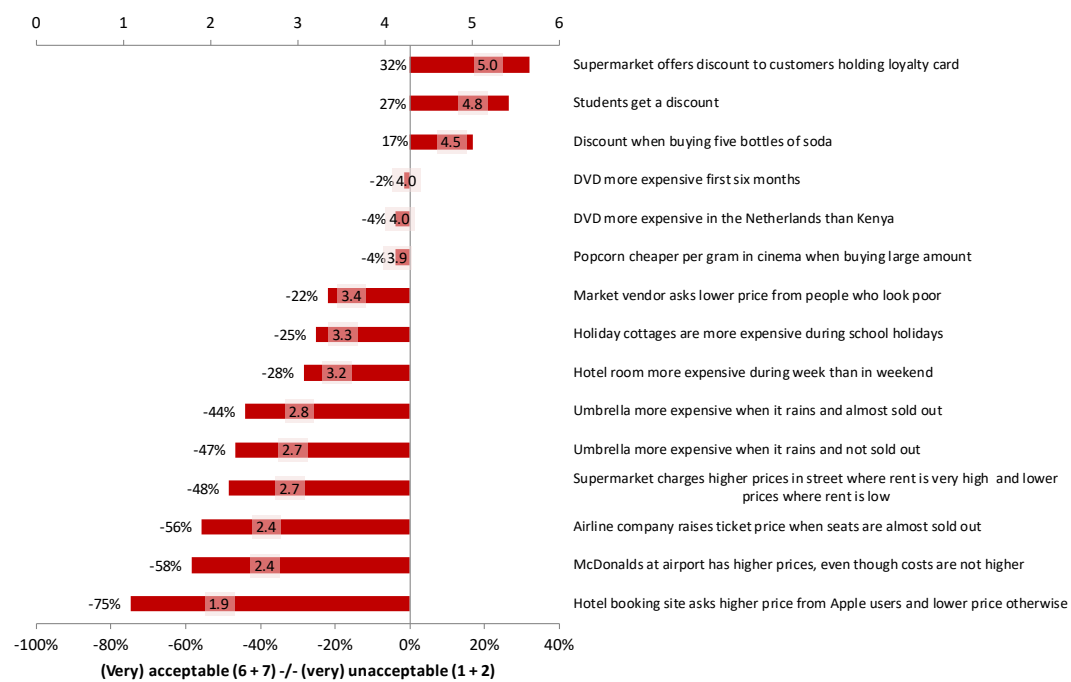

Figure 3 summarizes the answers to all fifteen questions by presenting the 'net acceptability'. We define net acceptability as the difference between the percentage that indicates 6 or 7 (acceptable or very acceptable) and the percentage that indicates 1 or 2 (unacceptable or very unacceptable). If 40% of respondents find an example (very) acceptable and 30% (very) unacceptable, the net acceptability is therefore 40% – 30% = 10%. In Figure 3, this net acceptability is sorted from high to low. The number within the bars gives the average of the scores: this ranges from 5.0 to 1.9 (4 is neutral) and strongly correlates with the net acceptability.

In general, Figure 3 illustrates that consumers reject many forms of price discrimination and dynamic pricing: for nine out of fifteen examples, the average score is below 4 and the net acceptability is well below 0%. Only three examples are regarded as predominantly acceptable: a supermarket offering discounts to customers with a loyalty card, a student discount and a quantity discount on bottles of soft drinks. Those surveyed reacted neutrally to three other examples, but found some examples unacceptable, even though such practices have been around for decades, such as airlines that raise their prices when the seats are almost sold out, and holiday homes that are more expensive during school holidays.

### 3.3 Analysis of the results: fairness, transparency and choice

These results show that a majority of the population finds price discrimination consistently unfair and unacceptable. Why do most consumers feel so uncomfortable with the practices of online price discrimination and dynamic pricing? Figure 2 shows that acceptance hardly depends on whether the price discrimination favours respondents themselves or others. There are also some (albeit inconclusive) trends in the answers to the second survey and the characteristics of the examples.[24] Respondents find volume discounts (second-degree price discrimination) relatively acceptable, while the picture for third-degree price discrimination is mixed, eg the supermarket loyalty cards are acceptable, but varying hotel prices based on the consumer's computer brands are not.

One key to understanding varying consumer sensitivity lies in the wording in which the price discrimination is framed. It is more acceptable if it is presented as a *discount* (or if the words 'cheaper' or 'lower price' are used) and less if it is presented as 'higher prices' or 'more expensive'. Behavioural economics has long shown that 'losses loom larger than gains'[25] and so the aversion to price discrimination can also be linked to an aversion to *loss* or *regret*.[26] Having said that, most consumers reject online price discrimination even when framed as a discount. Perhaps this is because many understand that there will be no gains without losses and that not receiving a discount resembles paying a premium.

Another key to understanding popular perception to price discrimination is that consumers appear to react more positively to price discrimination if it is more transparent. Student discounts and quantity discounts in supermarkets, for example, are generally transparent, whilst airline prices are opaque for most consumers. As mentioned, a majority of consumers worry that price adjustments may go unnoticed. On the other hand, discounts in the supermarket for customers with a loyalty card, the most accepted strategy in Figure 3, can be personalised and opaque; customers rarely see the size of the discount of other customers, although they might assume equality of treatment. By the same token, it is transparent that holiday homes are more expensive during school holidays, but consumers still do not like it.

A third key to making sense of consumer reactions to price discrimination is choice; a lack of real choice undermines the perceived legitimacy of price discrimination as, for example, 'trapped' customers behind the customs of an airport who are met with inflated prices for food and drinks while waiting to board their planes, parents with young children and holiday prices during the school vacations. In contrast, anyone can get a loyalty card from a supermarket to qualify for discounts, and someone can decide to go for a quantity discount and buy several bottles of soda or a large bucket of popcorn.

Finally, the relationship between respondents' responses and demographic factors suggests that price discrimination is most accepted by young, highly educated men in higher-income groups, while older, less educated women in lower-income groups are the least accepting.[27] It is likely that the link with education is driven by a better understanding of the underlying logic or even the positive effects that such strategies can have on allocation and welfare. The positive correlation between acceptance and income could be seen as unexpected, because higher income groups will in most cases be the 'victims' of price discrimination and pay higher prices. It is tempting to conclude that the perceived fairness transcends self-interest here. An alternative explanation is that it simply makes less of a difference to richer consumers, or that they benefit indirectly because without price discrimination – for instance for holiday cottages – they would more often be crowded out by less wealthy consumers.

## 4    The GDPR and online price discrimination

In Europe, there are no specific rules on price discrimination, and thus *prima facie* it is allowed. Contractual freedom implies that price discrimination is generally legitimate; economic actors are, in principle, free to decide with whom to contract and on what terms, including the price. A seller may choose a price for a product or service, and if a consumer consents to that price, there is no problem, at least from the perspective of contractual freedom.

However, online price discrimination whereby prices are adjusted to individual consumers generally invokes the operation of the GDPR, as such price discrimination typically involves the processing of personal data.[28] The GDPR tries to ensure that the processing of personal data happens fairly, lawfully, and transparently, and grants rights to individuals whose data is processed ('data subjects'), and imposes obligations on organisations that process personal data ('data controllers'; we also speak of 'companies' for ease of reading).[29] Independent Data Protection Authorities monitor compliance. The GDPR applies when 'personal data' is 'processed' which practically refers to everything that could be done with personal data.[30] The GDPR defines personal data as 'all information about an identified or identifiable natural person…'[31] which in turn is defined in this way:

an identifiable natural person is one who can be identified, directly or indirectly, in particular by means of an identifier such as a name, an identification number, location data, an online identifier or one or more elements characterising the physical, physiological, genetic, mental, economic, cultural or social identity of that natural person.[32]

The definition of personal data is thus broad in scope and includes data that can be used for online price discrimination, such as tracking cookies, IP addresses, and similar identifiers.[33]

For example, suppose Alice is a regular customer of an ecommerce site, and logs in with her email address. The online store sees that Alice uses an expensive smartphone, and also knows that she often buys the most expensive brands. Thus, the store concludes that Alice is not price conscious, and charges Alice 10% extra on all purchases. The store recognizes Alice when she logs in with her email address and a password. Her email address is personal data, because Alice can be identified from her email address. The GDPR's processing definition applies to recognising someone from her email address and adjusting the prices for that person. There is also Bob who is not a customer yet, but the online store recognizes him by means of a tracking cookie that a partner company has placed on his computer, that collects (personal) data about Bob's browsing activities. The provider gives Bob a 5% discount, without telling him, in the hope that he will become a customer.

In both cases the online store processes personal data, which invokes the operation of the GDPR. The fact that the GDPR applies does not mean that the use of personal data is *per se* prohibited, but rather that an organisation, such as the online seller, must meet the GDPR's requirements, which in fact reflect some of the consumer expectations identified above.

### 4.1 Transparency obligations resulting from the GDPR

The principle of transparency is one of the main principles of the GDPR.[34] The GDPR provides a detailed list of the information to be disclosed. For example, a company must provide information about its identity and the 'purposes of the processing' and must provide more information when necessary to ensure fair processing.[35] Thus a company must inform its customers if it processes personal data to personalise prices. Transparency about price discrimination could reduce the current information asymmetry and would allow consumers to choose web stores that do not personalise prices (provided there is sufficient competition), or delete their cookies if that gives them a better deal.

For that very reason, it is likely that companies prefer not to tell customers about personalized prices, especially if they charge higher prices, considering the empirical evidence presented in the last section, which showed that most consumers find online price discrimination unfair and reject it. Hence, consumers could be expected to react negatively and look for 'fairer' prices elsewhere. Currently there does not seem to be any website that tells their customers that they are personalizing prices. Perhaps companies that fall within the ambit of the GDPR do not personalise their prices, or perhaps some do but fail to comply with the GDPR.

Alternatively, some companies may try to comply with the GDPR with a sentence such as: 'we use personal data to offer our customers better personalized services.' However, the GDPR requires companies to formulate and disclose a 'specified' and 'explicit' purpose for the collection of data and thus a general statement as above is unlikely to meet that requirement.[36] European Data Protection Authorities have emphasized that 'a purpose that is vague or general, such as for instance "improving users' experience" [or]"marketing purposes"… will – without more detail – usually not meet the criteria of being "specific".'[37] In short, the GDPR requires transparency and obliges companies to inform their customers about the personalisation of prices explicitly and clearly.

**4.2   The GDPR and the requirement for legal basis**

Another core principle of data protection legislation is the requirement that companies that use personal data must have a legal basis for doing so. The GDPR provides an exhaustive list of six possible principles for the processing of personal data. The three legal bases that are most relevant to the private sector are consent, necessity for contractual performance and legitimate interest.[38]

First, a company may have a basis for processing personal data if the data subject has given consent to the processing of his or her personal data for one or more specific purposes.[39] The requirements for valid consent are strict. The GDPR requires a 'freely given, specific, informed and unambiguous indication of the data subject's wishes by which he or she, by a statement or by a clear affirmative action, signifies agreement to the processing of personal data relating to him or her.'[40] However, given the aversion of most consumers to price discrimination, it seems unlikely that many would give their consent. It is possible, however, that many users would click 'I agree' without reading the privacy notice.[41]

A second possible basis is that the 'processing is necessary for the performance of a contract to which the data subject is party or in order to take steps at the request of the data subject prior to entering into a contract.'[42] For example, if someone buys a book online, his or her address, which is personal data, would be required to deliver the book. The use of the address would be 'necessary for the performance of the contract', ie the delivery, but not for personalised pricing. 'Necessity' has a high threshold and is interpreted restrictively,[43] and thus the fact that a company might consider using personal data useful or profitable to price discriminate, does not make such use 'necessary.'[44]

Third, the data controller's 'legitimate interest' can sometimes provide a valid basis for using personal data in the absence of consent. A controller can rely on this legal basis when personal data usage is necessary for the controller's 'legitimate interests' and those interests are not 'overridden by the interests or fundamental rights and freedoms of the data subject which require protection of personal data'.[45] This legal basis seeks to strike a fair balance between the company's interests (ie. profit) and the consumer's interests (eg privacy, or low or fair prices). The legal basis is appropriate for standard business practices, such as marketing one's own products to existing customers. However, the legal basis does not seem appropriate for price discrimination, since this can hardly considered 'necessary.' Moreover, in the case of price discrimination, the consumer's interests would generally weigh heavier than the controller's interests. Indeed, Data Protection Authorities have declared that price discrimination is an example of a practice that cannot usually be based on the provider's legitimate interests.[46]

In conclusion, consumer's consent appears to be the only available legal basis for price discrimination based on personal data.[47] Yet, as mentioned above and in light of the survey results, consumers are unlikely to give their consent to such practices. If, regardless of this requirement of consent, online providers use personal data for price discrimination, Data Protection Authorities could enforce the GDPR and impose serious fines, of up to 20 million Euro or up to 4% of a company's worldwide turnover.[48]

## 5 Conclusion

While personalised pricing is far from a novel tool for sellers to appeal to different audiences, the explosion of online shopping has enabled much more precise price discrimination. Online stores can offer *each* consumer a different price based on detailed histories and profiles (unless those consumers take often elaborate and conscious steps to minimise their online footprint). Such new forms of price discrimination exacerbate concerns regarding the fairness, morality, and regulation of price discrimination.

One regulation that seems to offer a potential lever against the power that price discrimination grants to sellers, is data protection law. The GDPR applies when companies personalise prices and obliges them to inform customers about the purpose of processing their personal data and ask for their consent before using their personal data to personalise prices. The fact that we do not know of any companies requesting such permission may imply several things: either that websites are not yet engaging in price discrimination using personal data, or that they do so without complying with the GDPR on this point. If a Data Protection Authority suspects that a web store is personalizing prices, it could start an investigation, and impose fines for non-compliance with the GDPR. Such investigations are, however, slow, costly and cumbersome. Nevertheless, at least in principle, the GDPR seems to be the most useful legal instrument to regulate online price discrimination.

Alternatively, specific rules for online price discrimination could be introduced to address its various controversial issues. That would require an answer from the legislator to questions such as: is price discrimination fair if it leads to higher prices for richer customers, and lower prices for poorer ones? Is 'fairness' the appropriate yardstick for regulation? Is price discrimination only problematic if it leads to illegal discrimination, eg on the basis of ethnicity, or should some forms of price discrimination be banned altogether?

Our surveys suggest that, were it decided by consumers, online price discrimination would be banned. Consumers tend to be critical and suspicious of online price discrimination. Most consider online price discrimination unacceptable and unfair, and are in favour of a ban. When stores apply online price discrimination, most consumers think they should be informed about it. To many, the lack of transparency which often surrounds online price discrimination seems to violate the 'golden rule' as advocated by J.C. Penney, to treat others as you would like to be treated. And yet J.C. Penney went bankrupt, while dynamic pricing is rampant on the internet, and companies are continuing to experiment with different kinds of price discrimination.

---

[1] JC Penney Newsroom, "About JCPenney" (*JCPennyNewsroom.com*) <https://www.jcpnewsroom.com/about-company-info.html>.

[2] Parts of this chapter are based on J Poort and FJ Zuiderveen Borgesius, 'Does everyone have a price? Understanding people's attitude towards online and offline price discrimination' (2019) 1 Internet Policy Review 1; FJ Zuiderveen Borgesius and J Poort, 'Online Price Discrimination and EU Data Privacy Law' (2017) 40 Journal of Consumer Policy 347; FJ Zuiderveen Borgesius, 'Algorithmic decision-making, price discrimination, and European non-discrimination law' (2020) 31(3) European Business Law Review 401. The authors thank Claes de Vreese, Natali Helberger, Sophie Boerman, and Sanne Kruikemeijer for their input for the survey.

[3] The word 'identical' is crucial here, because it excludes price differences resulting from differences in the cost of serving different customers. Think for example of different shipping costs or different risk profiles in insurance and credit markets. Based on demographics or a person's past, he or she may be more likely to cause a traffic accident, fall ill, become unemployed or default on a loan. As a result, the cost of providing insurance or credit will differ. These cost differences justify price differences that most authors would not consider price discrimination. In fact, one could argue insuring different people entails different products. Different versions of a product sold with different margins also fall outside the scope of this chapter. An economically more profound definition of price discrimination by Stigler is 'the sale of two or more similar goods at prices which differ in relation to marginal costs'. GJ Stigler, *Theory of price* (4th edn, Macmillan 2003) 210. Under that definition, price differences which merely stem from cost differences do not constitute price discrimination. Versions may qualify as a result of the rather vague word 'similar'.

[4] See also Zuiderveen Borgesius (n 2).

[5] Inge Graef, 'Algorithms and Fairness: What Role for Competition Law in Targeting Price Discrimination Towards End Consumers?' (2018) 24(3) Columbia Journal of European Law 541.

[6] Tycho de Graaf, 'Consequences of Nullifying an Agreement on Account of Personalised Pricing' (2019) 8.5 Journal of European Consumer and Market Law 184.

[7] FJ Zuiderveen Borgesius, *Improving Privacy Protection in the Area of Behavioural Targeting* (Kluwer Law International 2015) <http://hdl.handle.net/11245/1.434236>.

[8] W Baker, M Marn and C Zawada, 'Price smarter on the net' (2001) 2 *Harvard Business Review* 122.***

[9] See for instance P Krugman, 'Reckonings; What Price Fairness?' *New York Times* (New York, 4 October 2000) <www.nytimes.com/2000/10/04/opinion/reckonings-what-price-fairness.html>.

[10] Amazon News Room, 'Amazon.com issues statement regarding random price testing' (27 September 2000) <http://phx.corporate-ir.net/phoenix.zhtml?c=176060&p=irol-newsArticle_PrintandID=502821>.